\documentclass[12pt]{spieman}  % 12pt font required by SPIE;
\usepackage{natbib}
\usepackage{graphicx}
\usepackage{hyperref}
\usepackage{amsmath}
\usepackage{mathtools}

\DeclareMathOperator{\sinc}{sinc}

\title{Electric Field Conjugation for Ground Based High-Contrast Imaging: Robustness Study and Tests with the Project 1640 Coronagraph}
\author[a]{Christopher T. Matthews}
\author[a]{Justin R. Crepp}
\author[b]{Gautam Vasisht}
\author[b]{Eric Cady}
\affil[a]{University of Notre Dame, Department of Physics, 225 Nieuwland Science Hall, Notre Dame, IN, USA, 46556}
\affil[b]{Jet Propulsion Laboratory, California Institute of Technology, 4800 Oak Grove Drive, Pasadena, CA, USA, 91109}
\begin{document} 
\maketitle

\newcommand{\be}{\begin{equation}}
\newcommand{\ee}{\end{equation}}
\newcommand{\Mj}{M$_{J}$}
\newcommand{\Ms}{M$_{\odot}$}
\newcommand{\Me}{M$_{\oplus}$}
\newcommand{\Mplanet}{M$_{planet}$}
\newcommand{\Rearth}{R$_{\oplus}$}
\newcommand{\LoD}{$\frac{\lambda}{D}~$}

\begin{abstract}
%The electric field conjugation (EFC) algorithm has shown promise for efficiently removing residual scattered starlight from high-contrast imaging measurements, both in numerical simulations and laboratory experiments. To prepare for the on-sky deployment of EFC using ground-based telescopes it is prudent to investigate the response of EFC to unaccounted for deviations from an assumed ideal optical model. We explore the linear nature of the algorithm by assessing its response to a range of unmodeled aberrations that are generally present in real optical systems.  We find that the algorithm is particularly sensitive to unresponsive deformable mirror (DM) actuators, misalignment of the Lyot stop, and misalignment of the focal plane mask. Vibrations and DM registration appear to be less of a concern when compared to actual values expected at the telescope. We quantify how accurately one must model these core coronagraph components to ensure successful EFC corrections in practice. We conclude that while the condition of the DM can ultimately limit achievable contrast, EFC may still be used to create dark holes that improve the sensitivity of high-contrast imaging observations. Our results have informed the development of a full EFC implementation using the Project 1640 coronagraph and hyperspectral imager at Palomar observatory.  While we focus on a specific instrument our results are applicable to the many coronagraphs that may be interested in employing EFC.
The electric field conjugation (EFC) algorithm has shown promise for removing scattered starlight from high-contrast imaging measurements, both in numerical simulations and laboratory experiments. To prepare for the deployment of EFC using ground-based telescopes we investigate the response of EFC to unaccounted for deviations from an ideal optical model. We explore the linear nature of the algorithm by assessing its response to a range of inaccuracies in the optical model generally present in real systems.  We find that the algorithm is particularly sensitive to unresponsive deformable mirror (DM) actuators, misalignment of the Lyot stop, and misalignment of the focal plane mask. Vibrations and DM registration appear to be less of a concern compared to values expected at the telescope. We quantify how accurately one must model these core coronagraph components to ensure successful EFC corrections. We conclude that while the condition of the DM can limit contrast, EFC may still be used to improve the sensitivity of high-contrast imaging observations. Our results have informed the development of a full EFC implementation using the Project 1640 coronagraph at Palomar observatory.  While focused on a specific instrument our results are applicable to the many coronagraphs that may be interested in employing EFC.
\end{abstract}

% Include a list of up to six keywords after the abstract

% Include email contact information for corresponding author
{\noindent \footnotesize\textbf{*}Christopher Matthews,  \linkable{cmatthew@alumni.nd.edu} }

\begin{spacing}{2}   % use double spacing for rest of manuscript

%\graphicspath{{./figures/}}

\section{Introduction}
The direct detection of exoplanets is limited by starlight scattered or diffracted from the surface of optical components. Complex wavefront phase and amplitude errors introduced by imperfections in the optical surface from coating, polishing, etc. within the science instrument manifest as bright speckles at the focal plane detector \citep{Borde_Truab}. Residual speckles from the on-axis star generally remain orders of magnitude brighter than even young, self-luminous gas giant planets; these speckles currently limit our ability to directly image faint sub-stellar companions at small angular separations, even when leveraging the largest ground-based telescopes at near infrared wavelengths \citep{crepp_11}. As the errors are generally introduced downstream from the adaptive optics (AO) system wavefront sensor (WFS), they cannot be sensed with traditional AO techniques. Instead, it is best to use the science instrument itself to measure non-common-path errors (as well as amplitude errors which the WFS is is not sensitive to), which can then be corrected using carefully calibrated commands sent to the AO system.

Various coronagraphic systems have been studied for space applications \citep{TPF,AFTA,JWST_corona,trauger_traub_07,exoC,HabEx}. Such missions have tight tolerances on allowable wavefront error (WFE) and stability in order to generate sufficient contrast to directly image faint companions, driving the requirements for coronagraphic hardware development. Proposed space missions have also motivated the development of techniques for removing stray starlight via focal plane wavefront sensing \citep{AFTA_EFC}. Computer simulations of the speckle nulling algorithm \citep{Speckle_nulling} and the electric field conjugation (EFC) algorithm \citep{Giveon} have shown promise for achieving contrast levels of $10^{-10}$ at close angular separations using internal occulters.  Other focal plane wavefront sensing methods, such as the Self-Coherent Camera technique, can measure the electric field at the science detector and provide significant contrast improvements but require minor modifications to the coronagraphic hardware \citep{SCC}.%Minor modifications to the coronagraph itself, such as the Self-Coherent Camera design, offer alternative methods for reducing quasi-static speckles in the focal plane \citep{SCC}.

Focal plane wavefront sensing algorithms have also been successfully implemented in a laboratory setting generating contrast levels as much as $6 \times 10^{-10}$ at 4 $\frac{\lambda}{D}$ using EFC \citep{Lawson_13_corona_review}. The dramatic improvement in contrast was achieved in part due to the extreme stability of the optical test-bench used \citep{HCIT}. Such ideal conditions well-emulate a space platform, but ground based instruments are subject to a number of environmental factors such as vibrations, large temperature variations, and mechanical flexure that can easily limit the ability to realize such gains.
    
The same observing techniques originally designed and tested for space missions are now being prepared for use with large-aperture, diffraction-limited telescopes  \citep{Groundbased_EFC,Eric_efc}. For example, speckle nulling has been successfully tested using the Subaru Coronagraphic Extreme AO (SCExAO) system on the Subaru telescope.    Martinache  et  al. found that speckle nulling could improve the contrast in the targeted half plane of the detector, reducing the standard deviation of speckle brightness in the DM control region by a factor of three \citep{SCExAO_spk_null} . Improved speckle stability also increased image quality and allowed for more effective post-processing with angular differential imaging (ADI \citep{ADI}). The Gemini Planet Imager (GPI) team has also tested speckle nulling \citep{GPI_spk_null}, finding that contrast levels could be improved from $5.7 \times 10^{-6}$ to $1.0 \times 10^{-6}$ when using a lab AO simulator. Speckle nulling has been tested at both Keck and Palomar as well\citep{Bottom_16_SN}. While speckle nulling is a robust technique, its efficiency is not optimal as it incorporates no knowledge of the optical model and must probe and compensate for sets of speckles sequentially rather than reconstructing the entire underlying electric field within the desired ``dark hole" (DH) region of the image plane in each iteration.    
    
In this paper, we investigate the behavior of the EFC algorithm. While mathematically more sophisticated and elegant than speckle-nulling, EFC is inherently dependent on the optical model used to describe the propagation of starlight from the telescope through the instrument to the detector \citep{Energy_minimization}. This work investigates the impact of model errors on the behavior of the EFC algorithm, the effect of aberrations on the raw dynamic range of coronagraphs has been studied elsewhere (e.x. \citep{low_order_aberations_paper}). By slightly modifying the model used to generate synthetic data while leaving the prescription used for the EFC calculations unchanged, one can assess how inaccuracies in the simulation of individual components affect overall performance. The goal of this study was to establish acceptable levels for common error sources which may be encountered in practice in the environment of an observatory. We found that while this technique is limited by imperfect knowledge of the optics and their alignment, EFC still represents a promising method for improving raw contrast levels achieved at ground-based observatories. 
    
We have also performed lab experiments demonstrating the algorithm's utility using the Project 1640 coronagraph and integral field spectrograph (IFS) at the Palomar Hale 200" telescope \citep{hinkley_11}. EFC corrections have been implemented previously at Palomar, even used to observe the bright star Vega \citep{Eric_efc}; however, the wavefront sensing portion of EFC (see below) was replaced by measurements using a Mach-Zehnder interferometric wavefront calibration (CAL) system, which was sensitive only to optics before the Lyot stop \citep{vasisht_14}. We have replaced this step in the procedure with a fully common-path technique that is sensitive to all components by using the science detector itself to measure the electric field. 
    
\section{Numerical Modeling of the P1640 Instrument}

We chose to study the Project 1640 (hereafter, P1640) coronagraph installed on the 200-inch Hale telescope at Palomar Observatory where research investigating the EFC algorithm and other techniques has been on-going \citep{hinkley_11,Eric_efc}. Our calculations employ physically accurate Fresnel propagation of starlight as it passes through a complete model of the instrument including all of the optics from the Hale telescope through the AO system and onto the P1640 IFU lenslet array. This approach provides the most accurate representation of the P1640 optical system available and, unlike Fraunhofer (far-field) calculations, captures contributions from optics located intermediate between pupil plane and focal planes. 
    
% Describe P1640 and simulation configuration
    % # of DMs, pupil only?
    % layout of system
    % lenslet array
    % IFS
    % Type of coronagraph
    
The P1640 coronagraph makes use of the PALM-3000 AO system, which contains a 66x66 actuator high-order DM and a 349 actuator low order DM in a woofer-tweeter configuration \citep{P3K}.  The high-order DM is conjugate to the pupil while the low-order DM is conjugate to a point 780 m above the telescope\citep{P3K}.  P1640 is an Apodized Pupil Lyot Coronagraph (APLC) design feeding into a 32 channel integral field spectrograph (IFS) operating between 1.05 - 1.75 $\mu$m with spectral resolution of R$\approx$30 \citep{hinkley_11}.  The IFS uses a 200x200 lenslet array, a prism for the dispersive element, and a Teledyne H2RG detector.  The system includes an additional internal interferometric wavefront sensor (CAL system) which measures the light immediately before the Lyot stop \citep{vasisht_14}.

We use the PROPER optical propagation routines developed by John Krist \citep{PROPER}, a code library that has been rigorously tested and proven to accurately represent diffraction in unfolded linear systems \citep{Krist_11}. A total of 27 optics were modeled in the analysis including the Hale telescope, PALM-3000 AO system \citep{P3K_14}, and P1640 coronagraph up to but not including the IFS lenslet array.  Our final resolution matches the 3.4 pixel per $\frac{\lambda}{D}$ that we measured for the $\lambda$ = 1.64 $\mu$m wavelength slices of the P1640 datacubes. For our simulations we assumed that the atmospheric turbulence was well corrected by the AO system and that the residual wavefront error was dominated by the quasi-static spatial errors.  As such we did not simulate the closed-loop behavior of the AO system, assuming that we were correcting the residual speckles remaining after AO correction and wavefront calibration using the CAL system.  We did not model temporal evolution of the error profile introduced by temperature variation or shifts in the pupil optics.
%The system was considered to be static in that closed-loop operation involving the correction of atmospheric turbulence was assumed to average down to levels below that of speckles. 

\begin{figure*}[!t]
\makebox[\textwidth][c]{\includegraphics[width=1.0\linewidth]{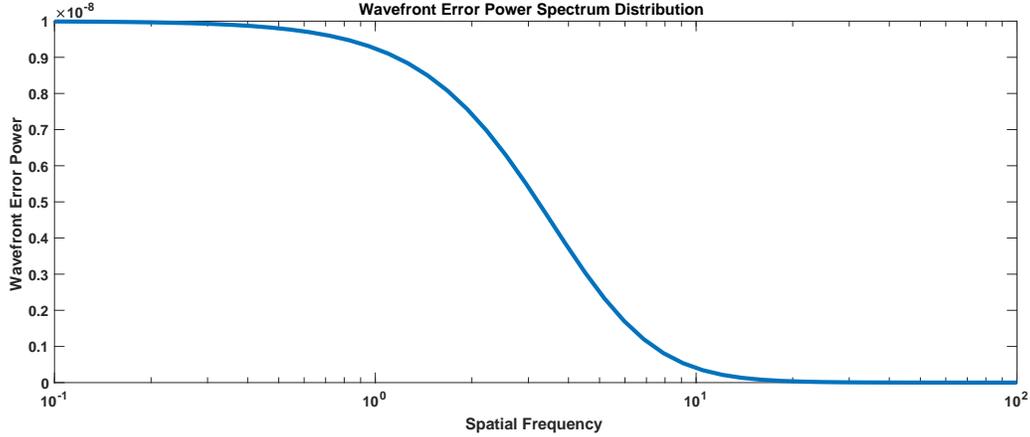}}
\caption{Power spectrum distribution function used to generate the initial wavefront error used in our simulations.}
\label{PROPER_psd}
\end{figure*}

Starting with the primary mirror, wavefront aberrations were added to optical components by generating phase-screens with PROPER's prop$\_$psd$\_$errormap routine which uses a power spectrum distribution (PSD) defined by:
\begin{equation}
%\begin{align*}
PSD(k) = \frac{a}{\left[ 1 + (\frac{k}{b})^{2} \right]^{\frac{c + 1}{2}}}
%\end{align*}
\end{equation}
where a = $1.0^{-8}$, b = $5.0$, c = $3.0$ for our simulations (See Fig. \ref{PROPER_psd}). The $\backslash$RMS flag was used so that the error map was normalized to have an rms value of 10 nm which we find best matches the rms phase error recorded with P1640 after corrections from the wavefront calibration unit (interferometer) have been applied.  The initial error was applied only as phase, the impact of initial amplitude errors will be addressed in future work.

\begin{figure*}[!t]
\makebox[\textwidth][c]{\includegraphics[width=1.0\linewidth]{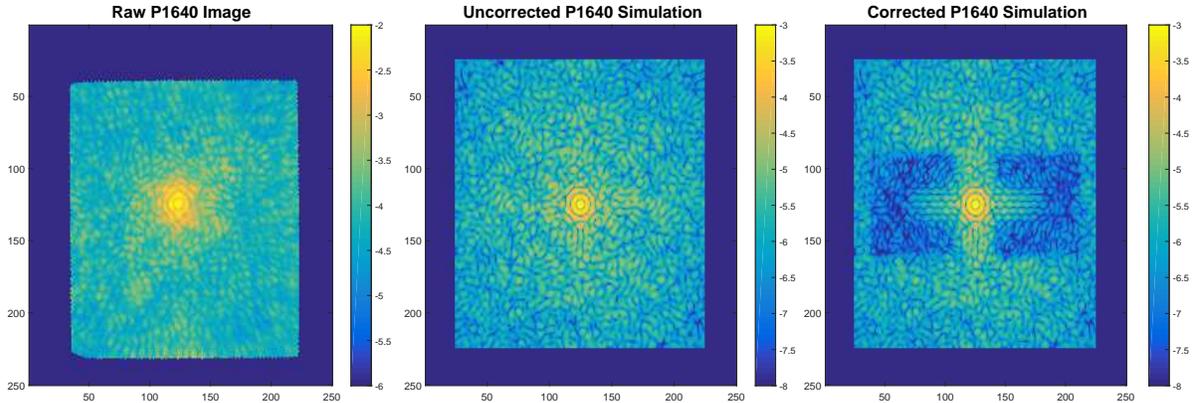}}
\caption{Comparison between the measured PSF taken from a P1640 datacube generated on 2012-06-15, a PSF simulated with our PROPER P1640 model, and the same PSF after 15 iterations of the EFC algorithm.  All three images are in units of contrast and plotted on a logarithmic scale.  Our simulations do not capture all of the low order error terms present in the P1640 image, nor do they include the additional speckle patterns introduced by the IFU optics and detector.  We also do not attempt to replicate the aberrations introduced by the current behavior of the DM (See discussion in Section 4).}
\label{p1640_sims_comp}
\end{figure*}

Fig. \ref{p1640_sims_comp}. shows an example comparing simulations to P1640 images. The 3d Fresnel calculations produce PSFs similar to those taken using the actual P1640 instrument, although they do not exactly match due to subtleties of the IFS, Hawaii-II HgCdTe detector, and datacube extraction pipeline. EFC consistently generates contrast improvements of 0.5-2.0 orders of magnitude within the DH control region in our numerical experiments. Some power remains at lower spatial frequencies due to residual diffraction from the spiders supporting the secondary mirror assembly; similar effects are present in P1640 datacubes though are often hidden in noise surrounding the coronagraphic mask.

EFC can effectively be divided into two core computational components. The first subroutine senses the complex electric field at the science detector using a series of ``probes" applied with the AO system's DM by employing a pair-wise phase diversity in order to reconstruct the electric field \citep{Giveon}. The second subroutine calculates the DM shape that minimizes the electric field in a given location based on the modeled impact that each DM actuator has on starlight at the detector.  This is done by building the Jacobian matrix which encodes the effect the DM has on the electric field in the detector focal plane.
    
In practice, building the Jacobian matrix that defines the influence of all individual actuators is done through simulations using an optical model.  However, deviations of the optical model from the actual configuration of the optics can be expected to impact the efficiency of EFC by requiring more iterations, limiting achievable contrast, or precluding convergence all together \citep{hcit_model_paper_2,hcit_model_paper_1}. Since it is not possible to have a completely accurate model of the instrument optics, as they change from hour to hour, night to night, and run to run, it is important to quantify the level of acceptable differences between the optical model and the physical optical train. 

The 66x66 actuator HODM provides a maximum control region corresponding to a half-plane dark hole (outer working angle) of 33 $\frac{\lambda}{D}$ wide. We chose to focus on a region centered at mid-spatial frequencies extending from 5 to 25 $\frac{\lambda}{D}$ in the horizontal direction and -10 to 10 $\frac{\lambda}{D}$ in the vertical direction.

\begin{figure*}[!t]
\makebox[\textwidth][c]{\includegraphics[width=0.8\linewidth]{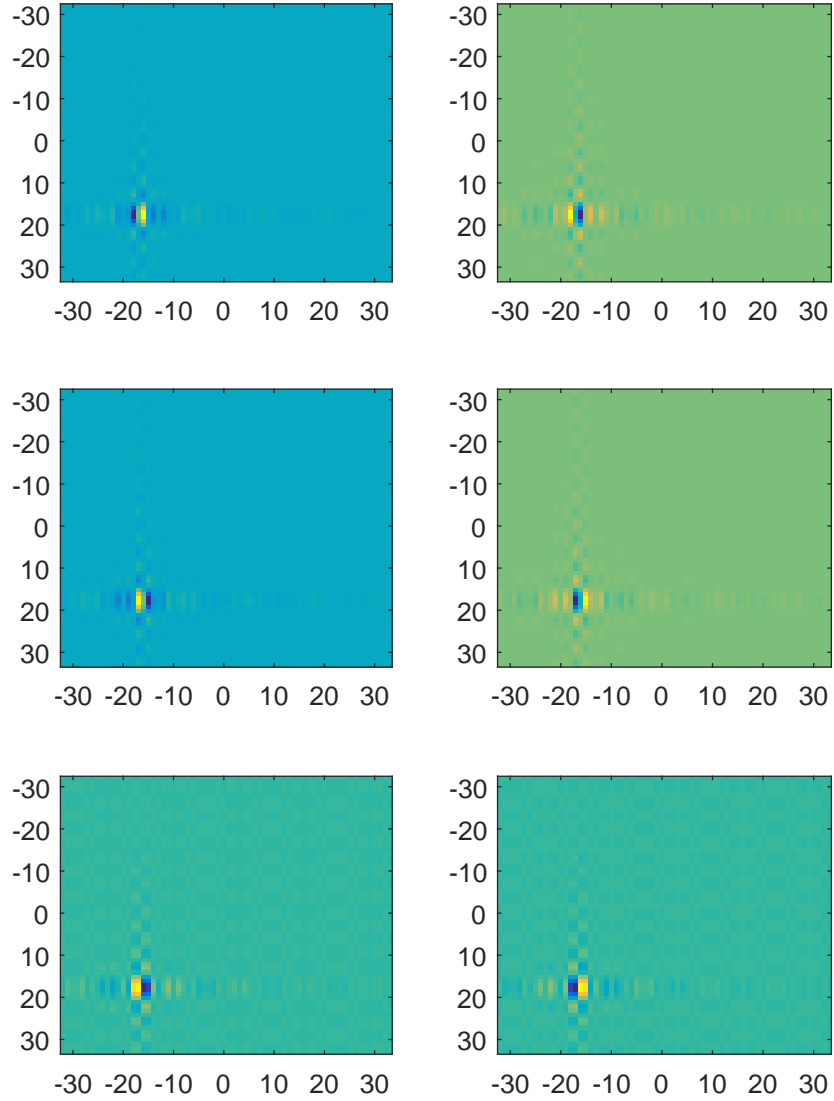}}
\caption{Profiles of the DM probes applied to the 66x66 DM in our Fresnel simulations, plotted on a linear scale and in units of nm of DM surface height.  The six probes form 3 $\pm$ pairs which was sufficient to reconstruct the electric field in our numerical experiments.  The probes were scaled to have a maximum/minimum amplitude of $\pm$ 20 nm.}
\label{probes_fig_WL}
\end{figure*}

The DM probe profiles used in our simulations were defined as:
\begin{equation}
\phi_j(u,v) = \sinc(f_{1} u)\sinc(f_{2} v)\sin(2\pi f_{3} u + \theta_{j}) 
\end{equation}
where $u,v$ are positions in the DM pupil plane and $f_i$ define spatial frequencies for the probe region \citep{Giveon}.  For the simulations presented here we used $f_{1} = f_{2} = 20$ and $f_{3} = 15$ to modulate a 20 x 20 $\frac{\lambda}{D}$ probe region centered at 15 $\frac{\lambda}{D}$.  This approach provides three sets of equal but opposite phase pairs for the reconstruction algorithm ($\theta_j = \pi/3, 2\pi/3, \pi, 4\pi/3, 5\pi/3, 2\pi$). The probes were scaled to have amplitudes of 20 nm.  Figure \ref{probes_fig_WL} shows the six DM probes used for the numerical experiments presented in this paper. 

We find that six probes allowed for reliable EFC convergence and correction within our computer simulations. In the lab however, experiments with P1640 performed in tandem with simulations have thus far required eight probes to produce dark holes due to the increased noise resulting from the P1640 datacube extraction process.  Additional probe images mitigate noise and insure that there is adequate modulation over the probe region to reconstruct the electric field in each pixel of the DH region.  We note that the probe profiles were offset from the center of the DM by 17 actuators in both directions to ensure that the peak of the probe pattern was not attenuated by the telescope central obstruction or spiders.

The PROPER framework places some practical restrictions on the focal plane resolution and DM actuator density that can be simulated. In particular, the second half of the EFC algorithm requires the inversion of a large matrix (the G-matrix) whose entries represent the effect each DM actuator has on the image plane electric field. The dimensions of this matrix are approximately given by (number of total pixels in focal plane) $\times$ (number of total actuators on the DM), which rapidly grew too memory intensive to handle. We chose to accurately represent the number of actuators on the 66 x 66 actuator PALM-3000 high-order DM (HODM) but did not model the low-order DM nor their closed-loop interaction. This decision is justified however by the fact that all EFC commands were only to the HODM to correct static errors. Further, to insure sufficient memory\footnote{Our simulations used a Dell Power Edge R815 Server with 128 GB of RAM} to perform the Moore-Penrose pseudo-inverse, the image plane resolution was limited to 250 x 250 pixels, which exactly matches the array size of a single wavelength slice ($\lambda =$1.64 $\mu$m) from the P1640 datacubes where the instrument is optimized. Resolution studies were performed showing that numerical noise limited us to a contrast floor of $\approx 10^{-8}$, sufficient to study ground-based coronagraphs.

Based on experience working with high-contrast imaging instruments in the past, we have decided to model the following errors as they may be expected to have an impact at the observatory:
    \begin{itemize}
    \item slowly changing registration of the DM
    \item misalignment of the Lyot stop 
    \item misalignment of the focal plane mask
    \item faulty DM actuators
    \item vibration from mechanical disturbances
    \end{itemize}
Additionally, telescopes with large primary mirrors generally contain substantial secondary obstructions and thus non-ideal pupils. It is not clear a priori how these various effects interact with one another within the context of EFC correction.  Table 1 summarizes the examples of model mismatch explored using EFC simulations and their error levels.  
    
\begin{table}[!t]
\centering
\begin{tabular}{lcc}
\hline
Model Error & Error Level \\
\hline
DM actuator response & 4 - 21 nm rms \\ 
Faulty actuators (full stroke) & 1 - 5 faulty actuators   \\
Faulty actuators (partial stroke) & 1 - 5 faulty actuators\\
Faulty actuators (zero stroke) & 1 - 5 faulty actuators \\
%DM stuck actuators (masked) & 1, 2, 3, 4, 5 masked actuators \\
%DM stuck actuators (masked, neighbors) & 1, 2, 3, 4, 5 actuators and their neighbors masked\\
Faulty actuators (floating) & 1 - 5 faulty actuators \\
DM registration & 1.0 - 5.0 $d_{DM}$ \\
%DM registration (with estimated correction) & 0.1, 0.2, 0.3, 0.4, 0.5 actuator spacing shifts\\
Mask alignment & 0.4 - 4.1 $\frac{\lambda}{D}$  \\
Lyot alignment & 0.01\% - 13.3\% of Lyot stop outer diameter\\
%Lyot rotation & 5nm \\
%Lyot dirty & 5nm \\
%Signal to Noise & Signal-to-noise of 100, 20, 10, 4\\
Vibration & 0.3 - 1.5 $\frac{\lambda}{D}$ rms \\
%Thermal & 2\%, 4\%, 6\%, 8\%, 10\% thermal background\\
%Planet 1 & 5nm \\
\hline
\end{tabular}
\caption{\label{Sim_param_p1640} Summary of the Fresnel simulations run to explore the effects of optical model inaccuracy on EFC performance.  Each potential model error was tested for 5 different levels inside the given range.  All of the simulations initialized with the same 10 nm rms phase error and were run for 15 EFC iterations.}
\end{table}

\section{Evaluation of the Impact of Optical Model Inaccuracies on EFC Performance}
\label{p1640_sim_results}

\begin{figure*}[!t]
\makebox[\textwidth][c]{\includegraphics[width=0.90\linewidth]{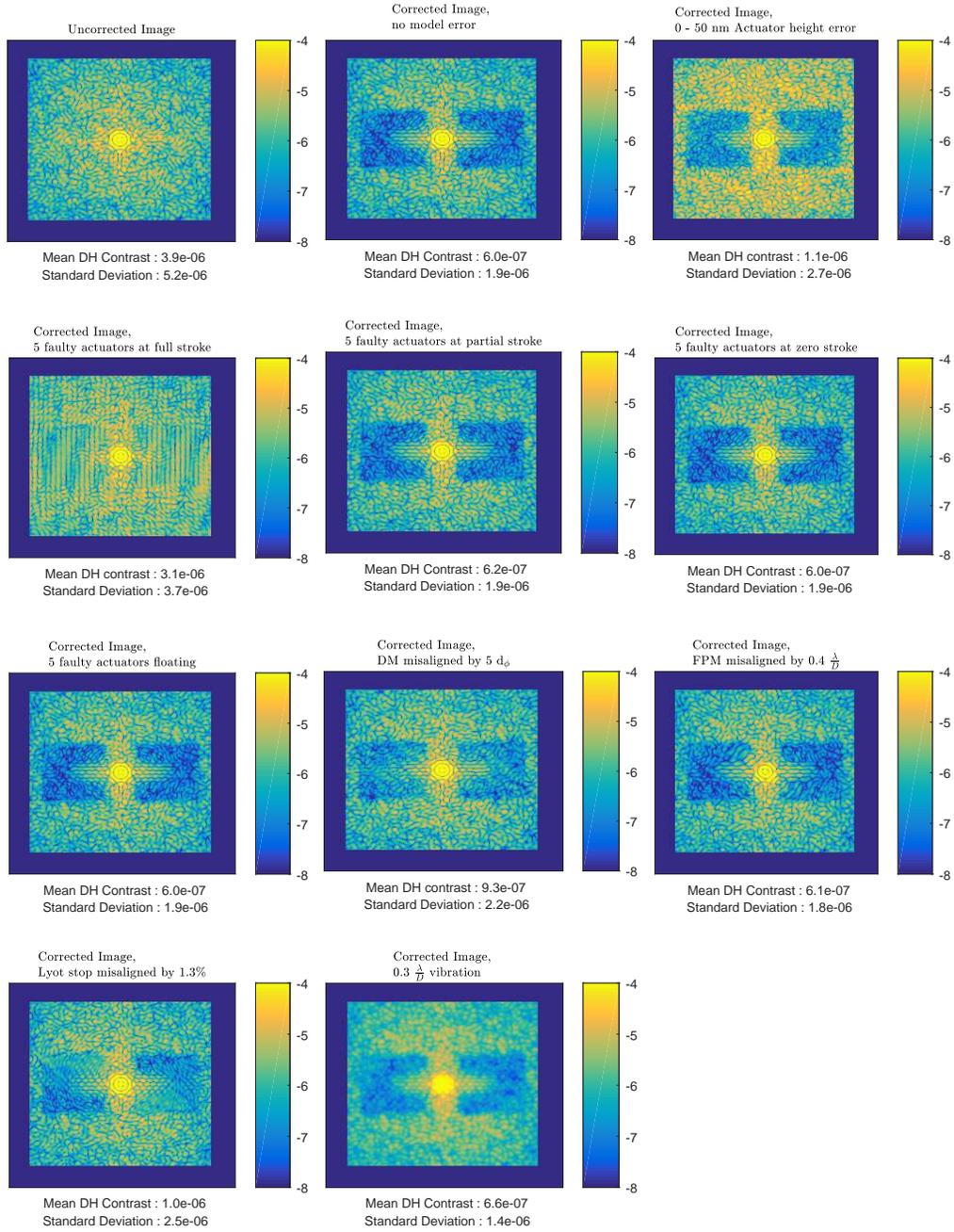}}
\caption{Comparison of example focal plane images of different model errors in our Fresnel simulations after 15 EFC iteration.  Examples are given for each of the model inaccuracies we simulated.  The top left image shows the uncorrected image with the initial 10 nm rms WFE.  The subsequent images are all after 15 EFC iterations.  All images are in units of contrast and plotted on a logarithmic scale.}
\label{DH_images}
\end{figure*}

Figures 5-13 show results for contrast versus angular separation for Fresnel simulations of potential deviations from the assumed optical model.  The 1-D contrast plots were generated by averaging the contrast in the y dimension for a strip along the x axis covering the DH region. Figure \ref{DH_images} shows examples of the focal plane images for each of the errors simulated.  The EFC algorithm shows a somewhat different response to each of the errors that were investigated which we discuss in turn. When compared to aberration levels expected at the observatory, the results may be used to establish effective tolerances when deciding whether to pursue EFC for a given high-contrast imaging system. 

\subsection{Uncertain Actuator Response}
\begin{figure*}
\makebox[\textwidth][c]{\includegraphics[width=1.0\linewidth]{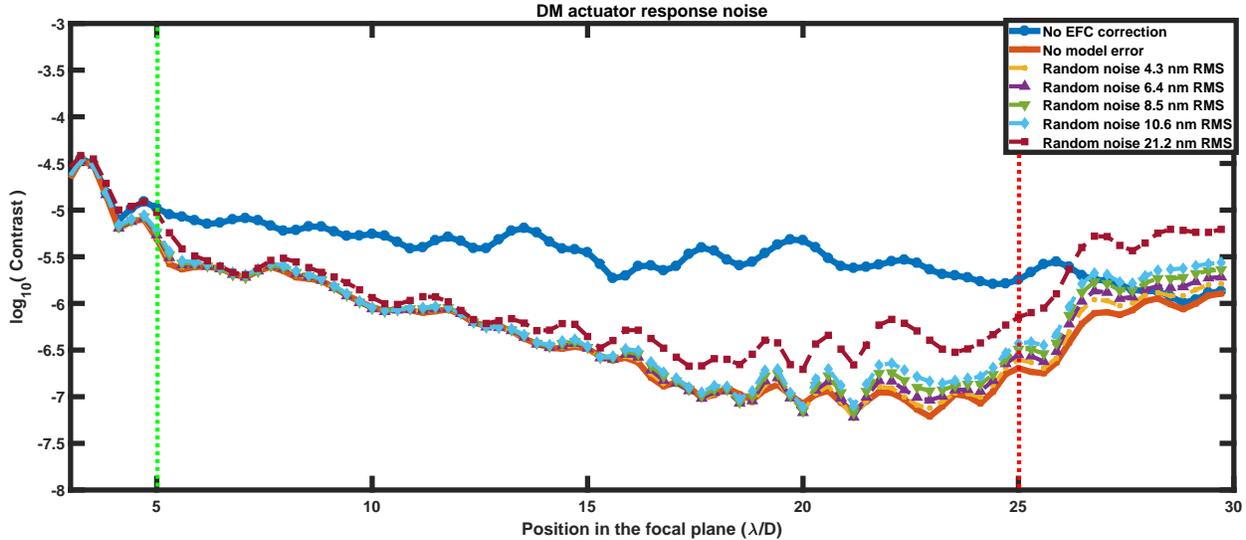}}
\caption{The effect of uncertain DM actuator response on the performance of the EFC algorithm when considering the full P1640 system. The injected noise is fit to a Gaussian random distribution with 0 mean and RMS values ranging from 4.3 to 21.2 nm.  Given the performance specifications of modern DMs this potential error source should not be significant.}
\label{DM_noise_p1640}
\end{figure*}

% 1. Actuator response. 
Actuator response inaccuracies were modeled using Gaussian distributed errors to simulate the effect of the DM creating unknown random phase errors (Fig. \ref{DM_noise_p1640}). Errors in the range of 4.3-21.2 nm rms were explored. We find that driving actuators to unknown (relative) phase locations at progressively larger amplitude minimally impacts the EFC algorithm compared to other error terms (see below). After 15 EFC iterations, this model error had essentially no effect on contrast until the injected noise became much larger than the $\approx 1$ nm precision of the P3k HODM. Still well below the wavelength of light, an uncertainty of 21.2 nm rms finally results in a noticeable degradation (a 10.9\% reduction in the median contrast inside the DH region). %As would be expected, inaccuracy DM actuator height primarily affects the higher spatial frequencies. 

\subsection{Faulty DM Actuators}
\begin{figure*}
\makebox[\textwidth][c]{\includegraphics[width=1.0\linewidth]{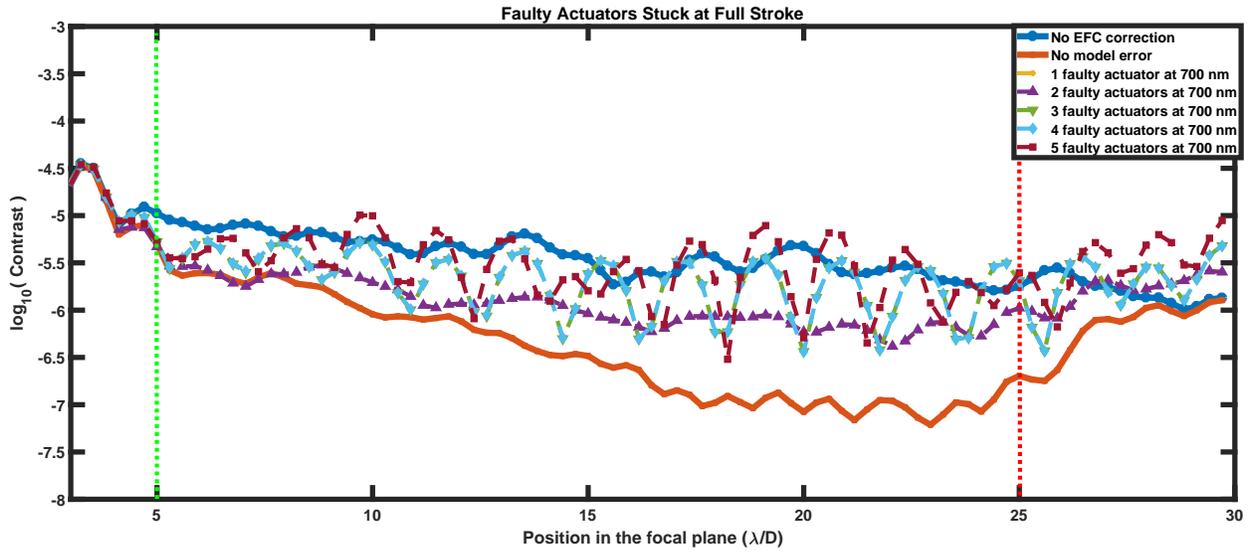}}
\caption{The results of our simulations exploring the effect of faulty actuators stuck at maximum stroke for simulations of the P1640 optical train.  Actuators stuck at such a large stroke (700 nm ~in these simulations) caused a decrease in the contrast improvement gained after 15 iterations of our EFC algorithm.  This is expected, as such a large phase change is comparable to the wavelength of the light we are simulating, and as such the linear approximation used in deriving the EFC algorithm is invalid.}
\label{DM_full_p1640}
\end{figure*}

\begin{figure*}
\makebox[\textwidth][c]{\includegraphics[width=1.0\linewidth]{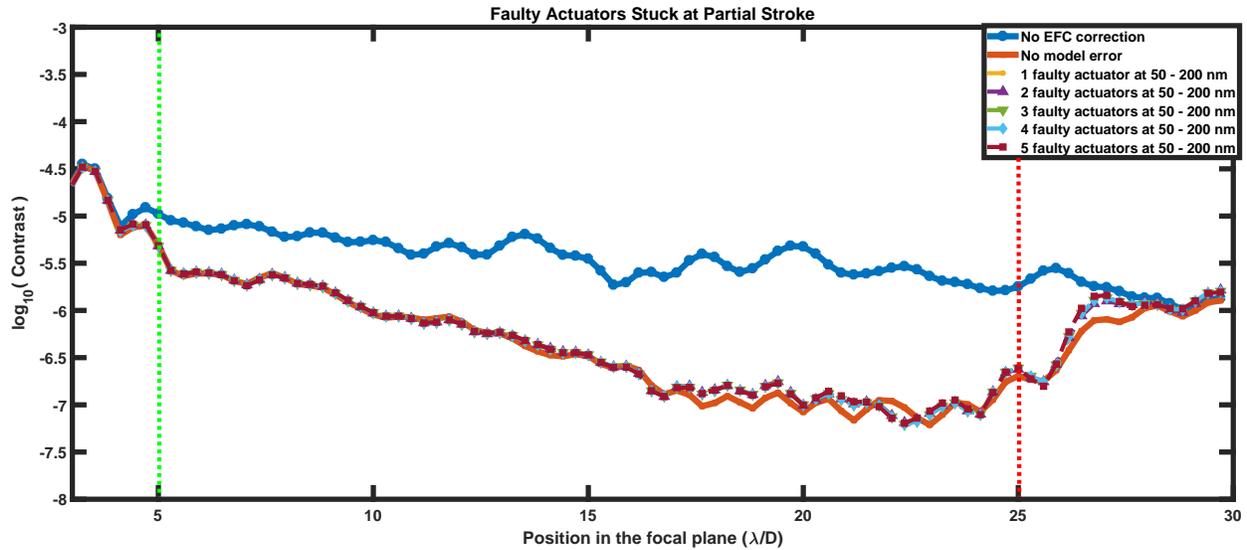}}
\caption{The results of the 3d simulations exploring the effect of faulty actuators stuck at a random partial stroke.  While some of these actuators are at a significant stroke (above 125 nm in these simulations), we did not see any significant drop in the effectiveness of our algorithm.  The damaged actuators in this experiment are not grouped, and all but one fall inside the transmissive area of the Lyot stop.  Please note that while all 5 curves are plotted in this figure, only the 4 and 5 damaged actuator simulations had an impact on the contrast.  The impact of 1-3 partially stuck actuators fell bellow the noise floor imposed by the numerical resolution of the simulations, indicating that this model error does not have a strong effect on the performance of the EFC code at least for the WFE regime currently of interest.}
\label{DM_part_p1640}
\end{figure*}

\begin{figure*}
\makebox[\textwidth][c]{\includegraphics[width=1.0\linewidth]{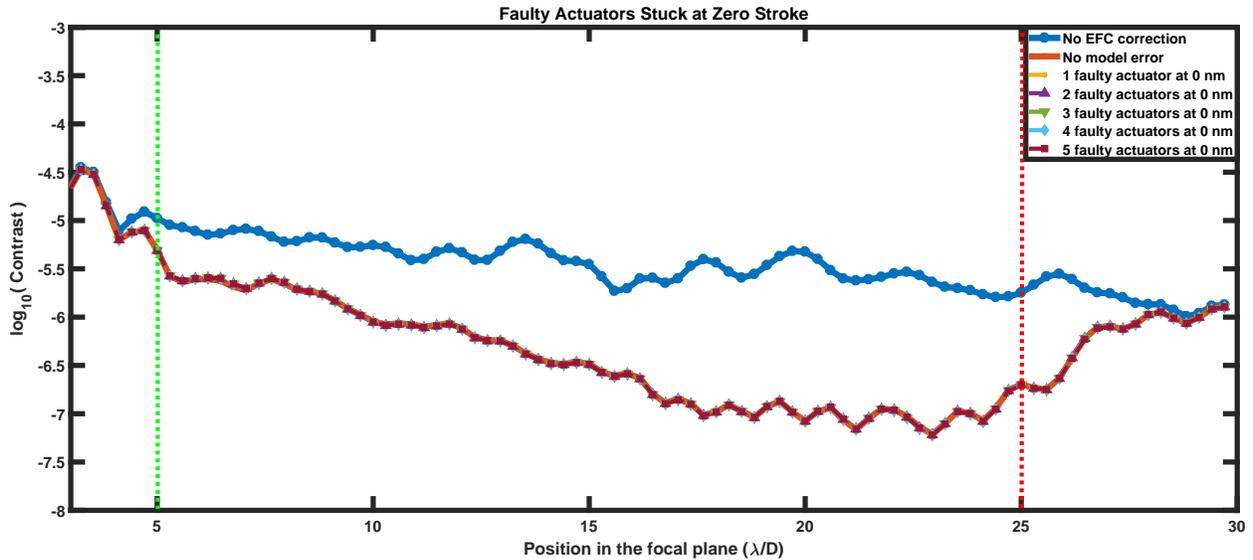}}
\caption{The results of our simulations exploring the effect of faulty actuators on the P3K HODM immovable at zero stroke.  Like the floating actuator situation these bad actuators had no effect on the contrast improvement in these simulations, implying that actuators stuck at zero stroke should not currently require consideration as EFC is implemented with P1640.  This is expected, the residual WFE in these simulations is small so losing 1-5 actuators does not preclude successful EFC correction.}
\label{DM_zero_p1640}
\end{figure*}

\begin{figure*}
\makebox[\textwidth][c]{\includegraphics[width=1.0\linewidth]{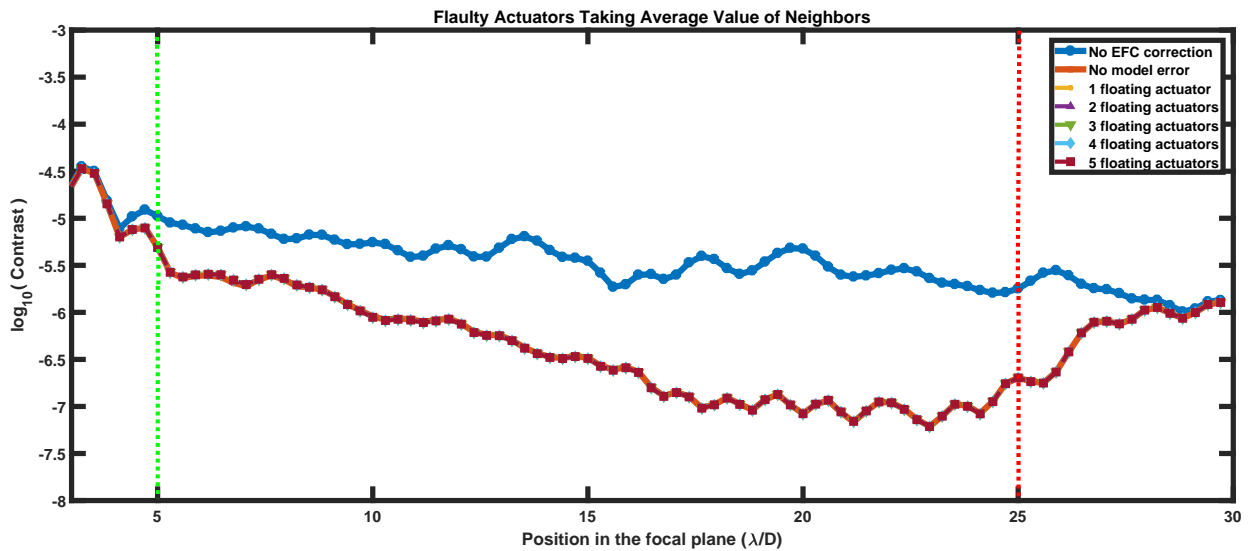}}
\caption{The results of our simulations exploring the effect of floating faulty actuators on the contrast improvement provided by my EFC code.  In this experiment floating actuators took the average value of both their horizontal and vertical neighbors.  we found that these damaged actuators produced no noticeable effect on the contrast achieved after 15 iterations of our EFC code.  Based on these results we conclude that floating actuators will not be a limiting factor when deploying EFC at Palomar.  While there are several groups of coupled actuators on the P3K HODM, their behavior is more complicated than the simple 'floating' prescription used in these simulations.  More detailed simulations of the HODM need to be run, but would require a better characterization of the state of the mirror than is currently available.}
\label{DM_float_p1640}
\end{figure*}

% 2. Damaged actutors. 
Damaged or partially inoperable actuators are a common problem with DM technology in general, particularly when related to the lifetime of hardware in an observatory environment \citep{GPI_dmgd_acts}. We simulate a variety of errors including actuators stuck at full stroke, partial stroke, zero stroke, and those that assume the height of their immediate neighbors, so called ``floating" actuators. We simulate a randomly distributed and ungrouped arrangement of 1-5 damaged actuators, as numbers much larger than this are known to create an immediate problem with dynamic range, EFC aside \citep{Damaged_acts}. For reference, the P3k HODM currently has two damaged actuators which are thought to be stuck at significant stroke. To help mitigate this effect, the P1640 Lyot stop is aligned such that one of the damaged actuators is occulted by material that also blocks starlight diffracted by the telescope spiders. Figures~\ref{DM_full_p1640}-\ref{DM_float_p1640} illustrate that DM actuators stuck at significant stroke (700 nm in this case) have a major impact on EFC's ability to converge (a single stuck actuator at full stroke reduced the improvement in median contrast in the DH region by 27.9\% and 5 stuck actuators reduced it by 88.4\%), while faulty actuators stuck at partial (50-200 nm) stroke, floating actuators, or those stuck at zero stroke have a minimal effect. In other words, while faulty actuators may ultimately limit the overall achievable contrast generated by a high-contrast imaging system, they do not necessary preclude the use of EFC. 

\subsection{DM Registration Error}
\begin{figure*}
\makebox[\textwidth][c]{\includegraphics[width=1.0\linewidth]{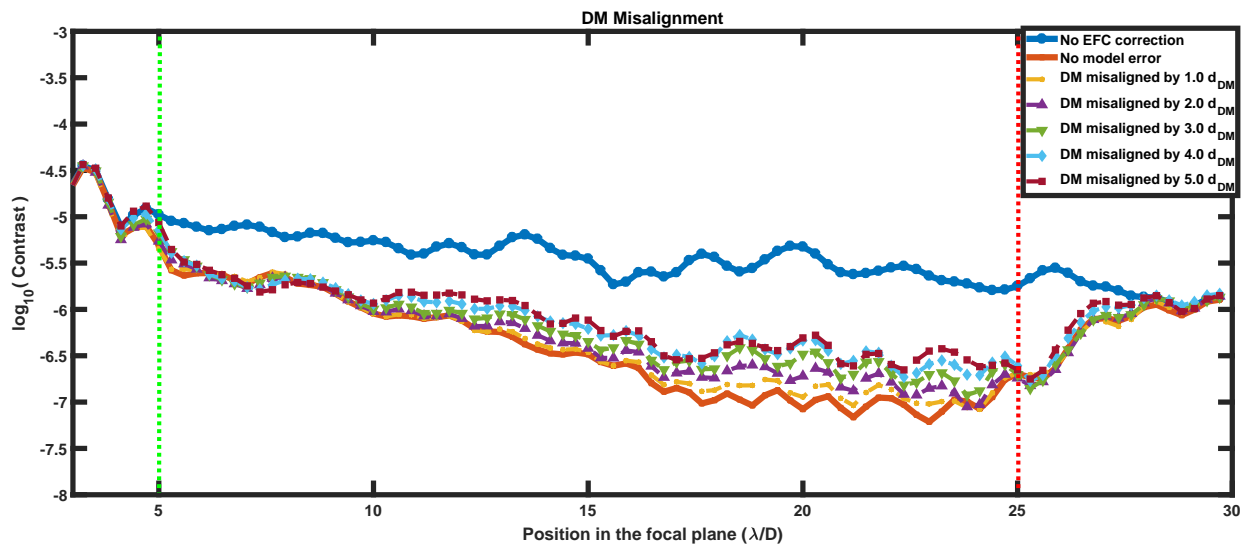}}
\caption{The effect of improper DM registration on the contrast improvement achieved by the EFC algorithm for the P1640 coronagraph with 10 nm initial RMS WFE.  Based on previous measurements from the CAL system, which measures the registration of the HODM relative to the CAL camera with each high order correction sequence, we expect the DM registration to vary on the order of 0.1 $d_{DM}$ over the course of the observation of a single target at the telescope, indicating that this error source should not be an issue when implementing the EFC code on sky at Palomar.}
\label{DM_reg_p1640}
\end{figure*}

% DM Registration. 
In addition to actuator response or faulty actuators, the DM itself could be globally misaligned with the optical pupil. We simulate registration errors, $d$, that range from 1-5 $d_{\rm DM}$ where $d_{\rm DM}$ is the rectilinear spacing of adjacent DM actuators (Fig.~\ref{DM_reg_p1640}).  The simulated DM was shifted in the x and y directions, we did not simulate translations in z (i.e. along the path of the beam). Simulations with $d < 1.0~d_{\rm DM}$ do not produce any noticeable impact on EFC's ability to generate a DH, and misalignments of 5 $d_{\rm DM}$ reduced the median contrast by only 10.4\%. In the case of P3k, the AO system is capable of routine coarse and fine alignment values well inside of this tolerance, a natural hardware requirement of the original design \citep{P3K}. We find that DM misalignment influences higher spatial frequencies more-so than lower spatial frequencies, which is a qualitative validation of our numerical simulations. 

\subsection{Focal Plane Mask Misalignment}
\begin{figure*}
\makebox[\textwidth][c]{\includegraphics[width=1.0\linewidth]{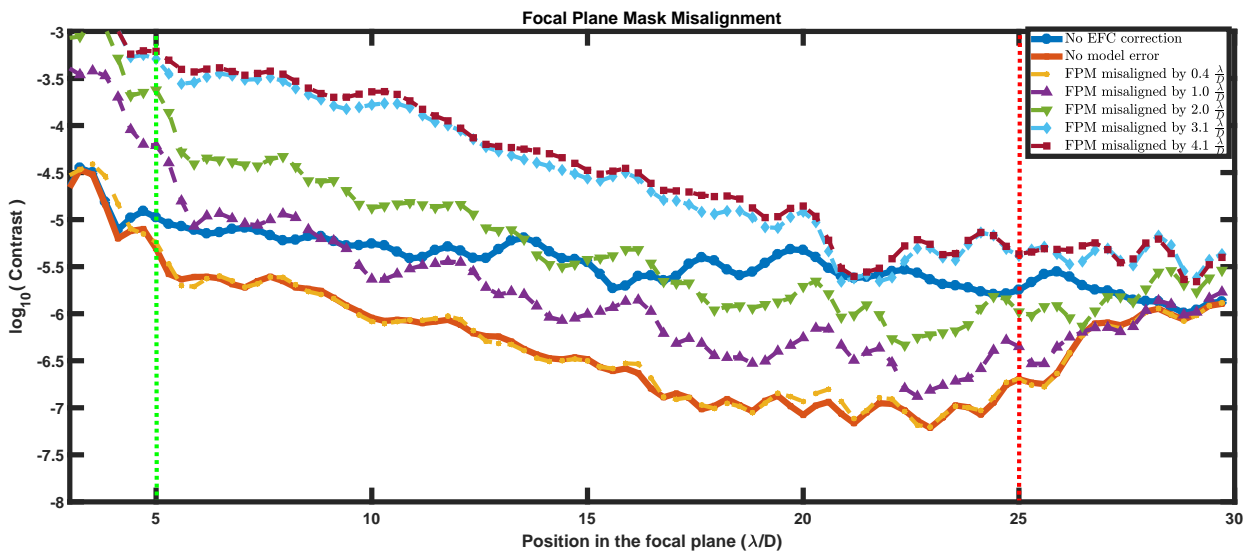}}
\caption{The 3d Fresnel simulations show that while misalignment of the focal plane mask can reduce the contrast improvement, this error source is manageable in the context of the P1640 coronagraph.  The mask within the P1640 coronagraph can be aligned with the beam to within 0.3 $\frac{\lambda}{D}$ thanks to the built in fine guidance sensor, so this error source should not pose a significant challenge to EFC implementation efforts.}
\label{mask_p1640}
\end{figure*}

% Focal plane mask. 
%Translations of the coronagraph focal plane mask (FPM) should begin to impact EFC when the error becomes comparable to $f \lambda / D$. P1640's focal plane mask has a diameter of 1322 $\mu$m, or 5.6 $f \lambda/D$.
The alignment of the coronagraph focal plane mask (FPM) is vital in maintaining the raw contrast provided by the instrument.  Results exploring FPM translation errors that range from 0.4 to 4.1 $\frac{\lambda}{D}$ are shown in Fig.~\ref{mask_p1640}. We find that shifts of the FPM greater than 0.4 $\frac{\lambda}{D}$ begin to impact EFC's performance in our simulations, with shifts of 2.0 $\frac{\lambda}{D}$ causing a 60.5\% reduction in median contrast improvement and larger shifts degrading the contrast compared to the uncorrected state. In practice, the P1640 coronagraphic mask may be aligned and maintained to within 75 $\mu$m (0.3 $\frac{\lambda}{D}$) using the quadcell fine guidance sensor behind the focal plane mask and the interferometric wavefront calibration unit. Thus, although the contrast degradation due to FPM translation is potentially large, because it governs starlight transmission in a focal plane, its effects on EFC can be easily mitigated with a well-calibrated coronagraph. 

\subsection{Lyot Stop Misalignment}
\begin{figure*}
\makebox[\textwidth][c]{\includegraphics[width=1.0\linewidth]{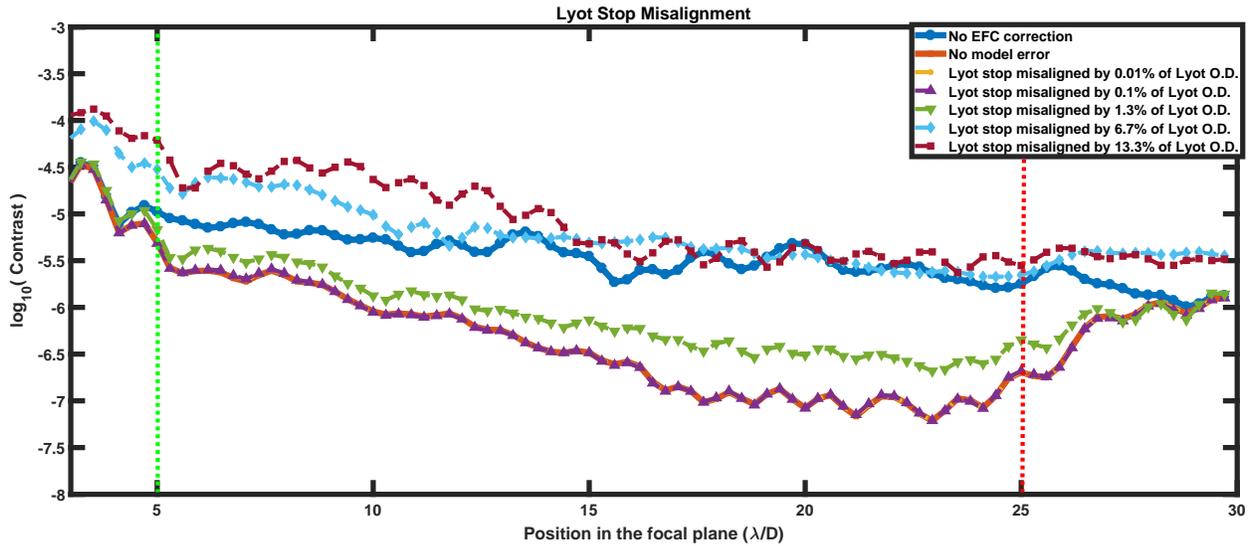}}
\caption{Our simulations show a limited dependence on the alignment of the Lyot stop when using the complete P1640 optical model.  The P1640 alignment procedure consistently centers the Lyot stop to within 1.3\% of the Lyot stop diameter, so while this potential model error may reduce the overall contrast improvement inside the DH  it should still be possible to achieve as much as an order of magnitude improvement in contrast when applying our EFC code at the telescope.}
\label{Lyot_p1640}
\end{figure*}

% Lyot stop. 
The Lyot stop works in tandem with the FPM to control diffracted starlight. As such, sufficiently large misalignments of the Lyot stop will degrade contrast and likely also prevent the EFC algorithm from working properly.  The Lyot stop in the P1640 coronagraph is 4\% undersized from the telescope pupil (2\% undersized from the apodizer) \citep{hinkley_11}. Results exploring unmodelled Lyot stop translations that range from 0.01 - 13.3\% of the Lyot stop outer diameter are shown in Fig.~\ref{Lyot_p1640}. Misalignments of 1.3\% result in a 12\% reduction in median contrast improvement after 15 iterations with larger shifts of the Lyot stop degrading contrast compared to the uncorrected image. For comparison, the P1640 Lyot stop may be repeatably aligned with the optical axis to within 1.3\% of the Lyot stop outer diameter following standard calibration procedures that involve imaging the pupil plane directly onto the detector. 

\subsection{Vibration}
\begin{figure*}
\makebox[\textwidth][c]{\includegraphics[width=1.0\linewidth]{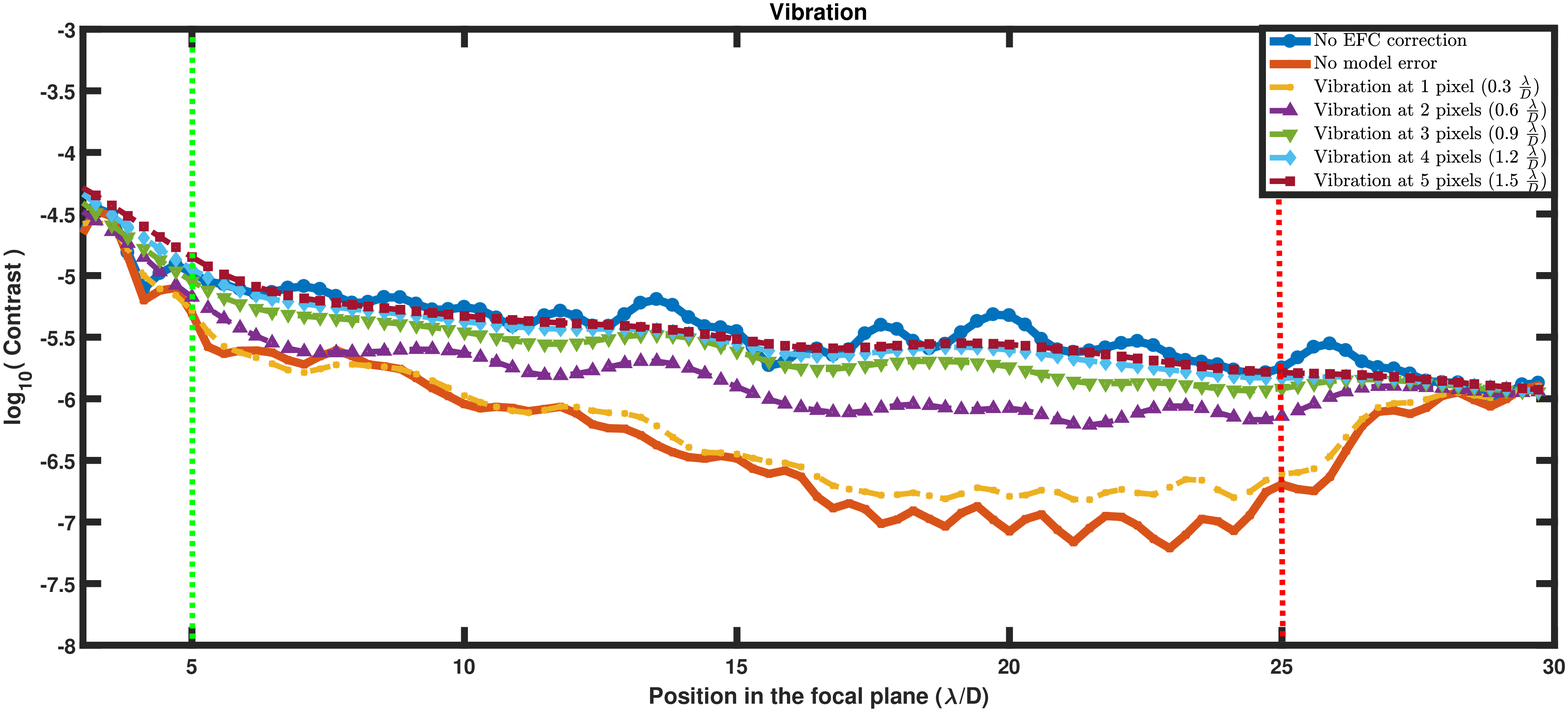}}
\caption{The effect of vibration on contrast improvement produced by EFC iterations show a very weak impact from vibration until the amplitude of the vibration grow larger than 1 pixel (0.3 $\frac{\lambda}{D}$).  We found that the degradation in contrast is very small for the levels of vibration which have been previously observed during P1640 operations.}
\label{vibe_p1640}
\end{figure*}

% Vibrations. 
Cryo-coolers or liquid nitrogen boil-off can introduce vibrations that spatially smear the image of the star at the detector, potentially impacting the efficiency of EFC \citep{GPI_IFS_vib}. Although vibrations are not noticed qualitatively with the P1640 system, the wavelength calibration unit uses a cryo-cooler and the IFS uses liquid nitrogen, so we decided to study the effect of spatial smearing by convolving the detector focal plane image with a Gaussian kernel that degrades speckle ``visibility."  We note that by using a Gaussian kernel we are assuming that the vibrations are isotropic, which is often not the case in practice.  As such our simulations only provide an upper bound on this model error.  Results exploring vibration levels that range from 0.3 to 1.5 $\frac{\lambda}{D}$, or 1 - 5 P1640 pixels, are shown in Fig.~\ref{vibe_p1640}. We find that contrast levels governed by the EFC algorithm become significant for vibration amplitudes of 1 pixel which reduces the median contrast improvement by 4.9\%.  Vibrations with amplitudes of 4 pixels or more effectively eliminate the contrast improvement from the EFC iterations. This result is consistent within the theoretical framework in which EFC operates on a pixel-by-pixel basis to clear out a DH. 
    
Table 2 summarizes the level of each simulated model error which resulted in a final contrast less than $1\times10^{-6}$ in the central 10 $\frac{\lambda}{D}$ of the DH region and compares those limits with the expected accuracy for the true instrument.  Given the results of our numerical simulations, which indicated that EFC may be a viable option for the P1640 coronagraph at Palomar, we decided to test our algorithm in with the actual instrument. A number of experiments were performed in 2015-2016 both in the Palomar AO lab as well as with P1640 mounted onto the Hale 200-inch telescope before and after science runs. These experiments helped us develop computer algorithms and scripts which interfaced with a number of hardware control systems and were used to establish the communication channels (between P1640, P3k, the wavelength calibration unit, and the telescope) required for EFC. 

\begin{table}[!h]
\label{accept_error} 
\centering
\begin{tabular}{lcc}
\hline
Model error & Limit based on P1640 simulations & Value expected at telescope \\
\hline
%Secondary obstruction & ? & ? $\frac{\lambda}{D}$ ? & ?\\
DM actuator response noise & 21 nm rms stroke & $<$ 1 nm \\
DM misalignment & 5.0 $d_{DM}$ & 0.1 $\Delta d_{DM}$ per hour\\
Actuators stuck at full stroke & 2 actuators & 2 dead actuators \\
Actuators stuck at partial stroke & 5 $+$ actuators&  unknown number of actuators \\
Actuators stuck at zero stroke & 5 $+$ actuators & 0 actuators \\
Floating actuators & 5 $+$ actuators & 0 actuators \\
Mask misalignment & 1.0 $\frac{\lambda}{D}$ & $<$ 0.3 $\frac{\lambda}{D}$ \\
Lyot stop misalignment & 2.6\% of Lyot stop O.D. & $<$ 1.3\% of Lyot stop O.D.\\
Vibration & 0.4 $\frac{\lambda}{D}$ vibration amplitude& $<$ 0.3 $\frac{\lambda}{D}$ vibration amplitude \\
\hline
\end{tabular}
\caption{Comparison between the results of our numerical experiments and the model accuracy expected with the P1640 instrument.  The simulation limits are the model error level which led to a final corrected contrast worse than 1 $\times10^{-6}$ average contrast in the central 10 $\frac{\lambda}{D}$ of the DH based on the 3d P1640 numerical experiments.  Our simulation results indicate that while the deterioration of the HODM may reduce the contrast improvement we can achieve, for realistic estimates of the inaccuracy in our optical model we still expect that approximately an order of magnitude improvement in contrast could potentially be achieved by applying the EFC algorithm.}
\end{table}

\section{Preliminary Demonstration of EFC Correction Using P1640} 

In preparation for the eventual on-sky deployment of our code we have performed a series of experiments performing EFC corrections with P1640 using the IFS.  These experiments were not intended to replicate the simulations presented in the previous section, but rather to assess the performance of the code under normal observatory conditions and to develop the software and procedures for EFC iterations using the instrument.  In this section we present an example of a successful set of EFC corrections using the IFS and discuss some of the practical lessons we have learned thus far.

% HODM situation. 
Based on our experience with the instrument we know that of the model errors we investigated with our simulations, only faulty actuators are present at Palomar at levels which may impede efficient EFC convergence.  The P3k HODM initially had several actuators with reduced stroke located within the telescope pupil (which subtends a circle projected across the DM surface), however these actuators are not of particular concern.  Since its introduction the surface of the Xinetics Inc. 66 $\times$ 66 actuator DM has evolved over time, developing several stuck actuators, several erratic actuators, as well as features that change from run to run (and sometimes night to night) based on external variables such as temperature and humidity levels. These idiosyncrasies, resulting from the individual reactions of the electrostrictive lead-magnesium niobate actuators and their electronics, were dealt with on a case-by-case basis, at times requiring that the influence of certain actuators be masked out or minimized with software if not by hardware. 

All of our lab experiments began with initializing the P3K AO system and aligning the P1640 optics on the P3K broadband stimulus source.  The source is a fiber-coupled Tungsten-Halogen lamp with a peak color temperature of 2800 K and an input current controller controlling the actual operating color temperature.  A single mode fiber relays the light to an off-axis parabola mirror upstream of the P3K system which collimates the light.  An additional aperture selects the central portion of the Gaussian beam to produce an approximately top-hat profile.  After alignment, low-order wavefront correction was applied using the internal Shack-Hartmann wavefront sensor followed by high-order wavefront correction with the CAL interferometric wavefront sensor.  These procedures routinely produced low-order phase error of $\approx$5 nm rms and high-order phase error of $\approx$10 nm rms with $~\approx$8\% residual amplitude error.  

Once wavefront correction was completed an initial set of images was taken with sine waves applied to the DM, both with the artificial source behind the focal plane mask and with the source offset from the mask.  These images were used to calculate a photometric normalization factor to convert the IFS output into units of normalized intensity which could be directly compared with the output of our numerical model.  EFC iterations were then performed.  First an unprobed image was taken, followed by a set of 8 images with phase diversity probes applied to the DM.  The extracted datacubes were then transferred to the EFC computer and used to reconstruct the electric field.  The electric field estimate was passed to the EFC correction subroutine which produced a set of DM commands which was then sent to the P3K system.  The correction was applied and used as the starting point for the next iteration.  With the current code each iteration took approximately 18 minutes, though this time could be reduced by better integrating the code into the P3K and P1640 control software.  The time required to take all 9 required datacubes was by far the longest, as we needed to use 90 sec exposures in order to avoid detector effects sometimes introduced by the changing clock state of the H2RG detector which would distort our electric field reconstruction.

\begin{figure}
\makebox[\textwidth][l]{\includegraphics[width=1.0\linewidth]{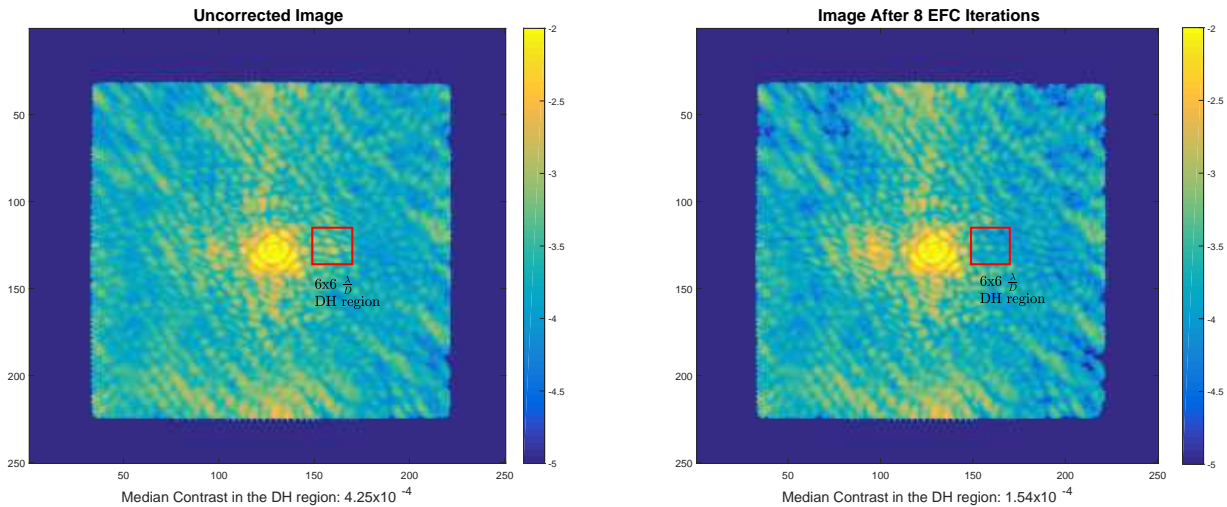}}
\caption{Comparison between uncorrected image (left) and an image taken after eight EFC iterations (right).  Both images are single wavelength slices from datacubes taken with P1640 mounted in the Palomar AO lab.  The red box indicates the 6x6 $\frac{\lambda}{D}$ target DH region. Eight iterations of our EFC code produced a 63.7$\%$ reduction in the median contrast inside the targeted region.}
\label{EFC_WL_DH}
\end{figure}

Based on the ultimate intended use for our EFC code, increasing signal to noise in observations of previously identified faint companions, we chose to focus our efforts on relatively small 6x6 $\frac{\lambda}{D}$ DH regions.  In order to ensure uniform modulation over the entire DH region we used a 14x14 $\frac{\lambda}{D}$ probe region with $\pm$ 100 nm probe amplitudes for the experiments presented in this section.  Figure \ref{EFC_WL_DH} shows the results of such an EFC experiment performed during an engineering run with the instrument in the Palomar AO lab.  After 8 iterations the EFC code successfully reduced the median contrast inside the DH region by a factor of 2.7 (a hardware issue precluded further iterations in this sequence).  Our experiments were focused on a single wavelength channel in the H band where the instrument is optimized, and we saw only limited improvement at other wavelengths.  We observed an improvement in final contrast in the 1.64 $\mu$m wavelength channel compared with previous EFC experiments using the electric field measured by the CAL interferometer rather than the phase-diversity based electric field reconstruction with IFS datacubes presented in this paper. 

%\begin{figure*}[!t]
%\makebox[\textwidth][c]{\includegraphics[width=0.75\linewidth]{probes_lab.eps}}
%\caption{Profiles of the DM probes applied to the 66x66 P3K HODM for EFC demonstration presented below, plotted on a linear scale and in units of nm of DM surface height.  The probes were scaled to have a maximum/minimum amplitude of $\pm$ 50 nm.}
%\label{probes_fig_lab}
%\end{figure*}

% Lab Results. 

%We also tested the speckle-nulling technique which does not depend on an accurate optical model of the coronagraph (Borde et al. 2006). As expected, speckle nulling was far less sensitive to changes in the performance of the AO system. Although more robust than EFC, we found that EFC was more efficient and generated deeper contrast levels. %No we didn't....

%Figure~\ref{EFC_WL_DH} shows results for a small, half-plane dark hole targeting a 6 x 6 $\lambda/D$ region. 

Our experiments have demonstrated that EFC can be used with high-contrast IFS instruments. Comparison with previous EFC experiments using P1640 indicate that while more complicated and time intensive reconstructing the electric field using measurements with the final science detector can provide more contrast in the DH region.  Our experiments thus far have also highlighted some of the technical challenges experienced in practice using a large ground-based telescope, particularly the sensitivity to detector effects and AO system behavior.  Further work is needed to improve the efficiency of our procedure before we can apply it on-sky.  Additional development is also necessary to improve the wavelength range of our DHs.  EFC can potentially improve the signal-to-noise ratio for spectra taken with IFS instruments, allowing for better characterization of exoplanets and other sub-stellar companions.
%Mention NCP errors and CAL and add redundancy to re enforce conclusions
%Mention increasing spectroscopy stn
%Mention small DHs?

%\begin{figure*}[!t]
%\makebox[\textwidth][c]{\includegraphics[width=0.8\linewidth]{probes_fig.eps}}
%\caption{Profiles of the probes used for the whitelight experiments.}
%\label{probes_fig_WL}
%\end{figure*}

\section{Summary \& Concluding Remarks}
The EFC algorithm could provide both theoretical and practical advantages over that of speckle-nulling for efficiently improving the dynamic range of high-contrast imaging instruments. While applications that mimic space environments have demonstrated the utility of EFC in lab experiments without a telescope \citep{Lawson_13_corona_review}, very little work has been done using existing ground-based telescopes. EFC is inherently a model-dependent algorithm in that convergence, the number of iterations, and performance noise floor depend upon previous knowledge and accuracy of the components that comprise the optical system. In practice, the geometric orientation of the telescope and instrument change in time when subject to mechanical flexure, thermal changes, and other effects that influence the alignment of an imaging system. 

We have developed a suite of numerical simulations that quantitatively address the tolerances that are imposed by inaccuracies in the optical model used by the EFC algorithm. Using the Palomar P3k AO system and P1640 high-contrast imager as an explicit example, we have used rigorous physical optics propagation to assess the response of the EFC algorithm to model mismatches. The code captures the effects of all optical components including those located in between pupil and imaging planes. Although more computationally intensive than Fraunhofer simulations, readily available multi-core computers can handle the inversion problem even when using large Jacobian matrices as required for a 66x66 actuator system such as P3k.

Perhaps not surprisingly, we find that EFC is most sensitive to faulty DM actuators stuck at significant stroke. However, actuators that ``float" and/or are stuck at partial stroke do not preclude the use of EFC. Following DM actuators, the next largest effects are misalignment of the Lyot stop and misalignment of the focal plane mask, but these errors are easily mitigated in practice using standard alignment procedures. Further yet down the priority list of error terms, registration of the DM with the telescope pupil and vibrations are, in a relative sense, much less of a concern when compared to values measured at the observatory. 

With the understanding that the tolerances for EFC were in an acceptable range, we then performed lab experiments to generate a dark-hole in the AO lab at Palomar using the full optical system illuminated by an optical fiber -- in preparation for an eventual on-sky demonstration at the telescope. We found that in practice the majority of the model-dependent aspects of an EFC iteration can be precomputed or simulated faster than the acquisition and extraction of datacubes from an IFS, thereby allowing us to integrate EFC into the P1640 observing sequence.

Previous attempts at EFC at Palomar using the same instrument employed an interferometer to directly measure non-common-path errors \citep{Eric_efc}, but were limited to accessing the beam before the light traversed the entire optical path. Our experiments used the P1640 IFS detector focal plane with fully reconstructed images to test EFC as it would ultimately be implemented on-sky. We find that EFC improved the average contrast in the DM control region by a factor of $\approx 2.5-2.7$, which is 25\% deeper than the results using an interferometer alone. While work still remains to examine the optimal number of DM probes, i.e. balancing time expended to record additional probe images against time saved through improved convergence and fewer iterations, our results indicate that EFC may indeed be used as (i) an efficient alternative to speckle-nulling, much like space applications, (ii) an independent method for validating the performance of wavefront calibration systems, and (iii) as a technique to enhance the sensitivity of high-contrast imaging instruments in general.

\section{Acknowledgements}
J.C. acknowledges support from the NASA Career program (NNX13AB03G). A portion of the research in this paper was carried out at the Jet Propulsion Laboratory, California Institute of Technology, under a contract with the National Aeronautics and Space Administration (NASA). We recognize the Potenziani family for their vision and support of Notre Dame astrophysics research. 

\bibliographystyle{apj}
\bibliography{biblio}

\end{spacing} 
\end{document}